\documentclass[useAMS,usenatbib]{mn2e}
\usepackage{graphicx,amssymb,hyperref}
\usepackage[fleqn]{amsmath}  
\usepackage{multirow}
\usepackage{graphicx} 
\usepackage{color}

\usepackage{graphicx,amssymb,multirow}
\usepackage[fleqn]{amsmath}

\def\simlt{\lower.5ex\hbox{$\; \buildrel < \over \sim \;$}}
\def\simgt{\lower.5ex\hbox{$\; \buildrel > \over \sim \;$}}

\def\ie{{\it i.e.}}

\newcommand{\be}{\begin{equation}}
\newcommand{\ee}{\end{equation}}
\newcommand{\ba}{\begin{eqnarray}}
\newcommand{\ea}{\end{eqnarray}}

\usepackage{graphicx} 

\title[Dark matter dynamics in Abell 3827]{Dark matter dynamics in Abell 3827:\\new data consistent with standard Cold Dark Matter} \author[R.\ Massey et al.]{Richard Massey$^{1,2}$\thanks{e-mail: {\tt r.j.massey@durham.ac.uk}},
David Harvey$^{3}$, 
Jori Liesenborgs$^{4}$, 
Johan Richard$^{5}$, \newauthor
Stuart Stach$^{2}$, 
Mark Swinbank$^{2}$,
Peter Taylor$^{6}$,
Liliya Williams$^{7}$, \newauthor 
Douglas Clowe$^{8}$, 
Fr\'ed\'eric Courbin$^{3}$, 
Alastair Edge$^{2}$,
Holger Israel$^{1,9}$, \newauthor
Mathilde Jauzac$^{1,2,10}$, 
R\'emy Joseph$^{3}$, 
Eric Jullo$^{11}$, 
Thomas D.\ Kitching$^{4}$, \newauthor
Adrienne Leonard$^{12}$,  
Julian Merten$^{13,14}$,
Daisuke Nagai$^{15}$,
James Nightingale$^{2}$, \newauthor
Andrew Robertson$^{1}$, 
Luis Javier Romualdez$^{16,17}$, 
Prasenjit Saha$^{18}$,
Renske Smit$^{19}$, \newauthor
Sut-Ieng Tam$^{2}$ \& 
Eric Tittley$^{20}$ \\
$^{1}$ Institute for Computational Cosmology, Durham University, South Road, Durham DH1 3LE, UK\\ 
$^{2}$ Centre for Extragalactic Astronomy, Durham University, South Road, Durham DH1 3LE, UK\\
$^{3}$ \'Ecole Polytechnique F\'ed\'erale de Lausanne, 51 Chemin des Maillettes, Observatoire de Sauverny, Versoix, CH-1290, Switzerland\\
$^{4}$ Expertisecentrum voor Digitale Media, Universiteit Hasselt, Wetenschapspark 2, B-3590, Diepenbeek, Belgium\\
$^{5}$ Univ Lyon, Univ Lyon1, Ens de Lyon, CNRS, Centre de Recherche Astrophysique de Lyon UMR5574, F-69230, Saint-Genis-Laval, France\\
$^{6}$ Mullard Space Science Laboratory, University College London, Holmbury St Mary, Dorking, Surrey RH5 6NT, UK\\
$^{7}$ School of Physics \& Astronomy, University of Minnesota, 116 Church Street SE, Minneapolis, MN 55455, USA\\
$^{8}$ Department of Physics and Astronomy, Ohio University, 251B Clippinger Labs, Athens, OH 45701, USA\\
$^{9}$ Fakult\"at f\"ur Physik, LMU M\"unchen, Scheinerstr. 1, 81679 M\"unchen, Germany\\
$^{10}$ Astrophysics and Cosmology Research Unit, University of KwaZulu-Natal, Durban 4041, South Africa\\
$^{11}$ Aix Marseille Universit\'e, CNRS, LAM (Laboratoire d'Astrophysique de Marseille), UMR 7326, 13388, Marseille, France\\
$^{12}$ FlightGlobal, RBI UK, World Business Center 2, Newall Road, Heathrow Airport, TW6 2SF UK\\
$^{13}$ Department of Physics, Oxford University, Keble Road, Oxford OX1 3RH, UK\\
$^{14}$ INAF, Osservatorio Astronomico di Bologna, via Pietro Gobetti 93/3, 40129 Bologna, Italy\\
$^{15}$ Department of Physics, Yale University, New Haven, CT 06520, USA\\
$^{16}$ Institute for Aerospace Studies, University of Toronto, Canada\\
$^{17}$ Centre for Advanced Instrumentation, Durham University, South Road, Durham DH1 3LE, UK\\
$^{18}$ Physik-Institut, University of Z\"urich, Winterthurerstrasse 190, 8057 Z\"urich, Switzerland\\
$^{19}$ Kavli Institute of Cosmology, Cambridge University, Madingley Road, Cambridge CB3 0HA, UK\\
$^{20}$ Royal Observatory, Blackford Hill, Edinburgh EH9 3HJ, UK
}
\begin{document}
\date{ Accepted ---. Received ---; in original form \today.}

\pagerange{\pageref{firstpage}--\pageref{lastpage}} \pubyear{2017}

\maketitle

\label{firstpage}

\begin{abstract}

We present integral field spectroscopy of galaxy cluster Abell\,3827, using ALMA and VLT/MUSE.
It reveals an unusual configuration of strong gravitational lensing in the cluster core, with at least seven lensed images of a single background spiral galaxy.
Lens modelling based on HST imaging had suggested that the dark matter associated with one of the cluster's central galaxies may be offset.
The new spectroscopic data enable better subtraction of foreground light, and better identification of multiple background images.
The inferred distribution of dark matter is consistent with being centered on the galaxies, as expected by $\Lambda$CDM. 
Each galaxy's dark matter also appears to be symmetric.
Whilst we do not find an offset between mass and light (suggestive of self-interacting dark matter) as previously reported, the numerical simulations that have been performed to calibrate Abell\,3827 indicate that offsets and asymmetry are still worth looking for in collisions with particular geometries.
Meanwhile, ALMA proves exceptionally useful for strong lens image identifications.

\end{abstract}

\begin{keywords}
dark matter --- 
astroparticle physics --- 
galaxies: clusters: individual: Abell~3827 --- 
gravitational lensing: strong
\vspace{-7mm}
\end{keywords}

\section{Introduction}

Determining the properties of dark matter has become a priority of astrophysics and particle physics.
In the standard $\Lambda$CDM cosmological model, dark matter has significant interactions with standard model particles through only the gravitational force \citep[e.g.][]{mrev,knerev}.
It therefore neither emits nor absorbs light, and appears invisible.
Nonetheless, over the course of cosmic history, dark matter's gravitational attraction assembled the Universe's large-scale structure, and governed the evolution of galaxies.
Dark matter has pulled together both ordinary and dark material into a series of collisions -- then eventual mergers -- between ever-larger structures \citep{illustris,eagle}.

Several particle physics theories of dark matter predict additional forces {\em between} dark matter particles, hidden entirely within the dark sector \citep{peterreview,atomicdm}.
The most direct way to measure these hypothesised forces is to observe the trajectory of dark matter during collisions with other dark matter.
In effect, astrophysical mergers can be treated as enormous particle colliders \citep{clo04,clo06,bra08,mer11,clo12,daw12,gas14,cho15,ng15,har15, jee16,gol16,gol17,kim17,mon17}. 
In simulated mergers assuming $\Lambda$CDM, the (non-interacting) dark matter remains tightly bound near stars \citep{sch15}. 
If dark-sector forces exist, simulations of mergers predict dark matter to temporarily lag behind stars, which serve as collisionless test particles \citep{ran08,mkn11,daw13,har14, kah14,rob17a,rob17b}.
In some simulations, the distribution of dark matter is also stretched into asymmetric tails \citep{kah14}.

Two properties of galaxy cluster Abell~3827 (RA=$22$h\,$01\arcmin$\,$49\farcs1$, Dec=$-59^\circ$\,$57\arcmin$\,$15\arcsec$, $z$=0.099, \citealt{depla07,car10,ws11}), make it uniquely interesting for studies of dark matter dynamics.
First, the cluster core contains four similarly-bright galaxies.
They must be undergoing a simultaneous, high speed merger, because this amount of substructure is unique: most clusters have reached a steady state with only a single Brightest Central Galaxy.
Second, directly behind the cluster core lies a spiral galaxy ($z\!=\!1.24145\pm0.00002$; \citealt{mas15}) that is rich in morphological structure.
The background spiral galaxy has been gravitationally lensed by the cluster, and its multiple images wrap around all four of the central galaxies. 
These images can be used to infer the spatial distribution of (dark plus stellar) mass in the cluster and its galaxies.

One of Abell~3827's central galaxies lies very close to a set of gravitationally lensed images, so the distribution of its mass is particularly well constrained.
Analysis of the gravitational lensing in optical imaging suggested that this galaxy's dark matter is offset by $1.62^{+0.47}_{-0.49}$\,kpc from its stars \citep{mas15}, and possibly asymmetric \citep{tay17}.
This could have been caused by a dark sector force with interaction cross-section $\sigma/m\simgt 1$cm$^2$/g, where $m$ is the (unknown) mass of the dark matter particle \citep{kah15}.
The most difficult part of this analysis was the identification of features in the faint, background spiral, right next to a very bright foreground galaxy (see Appendix~B in \citealt{mas15}).

In this paper, we present new Integral Field Unit (\ie\ 2D) spectroscopy of Abell~3827 at near-IR and millimetre wavelengths: where the foreground cluster is faint, but the background spiral galaxy remains bright.
We describe the new data in section~\ref{sec:data}.
We describe our analysis techniques in section~\ref{sec:analysis}, and reconstruct the spatial distribution of dark matter in section~\ref{sec:results}.
We discuss the consequences of our results in section~\ref{sec:conc}.
Throughout this paper, we adopt a cosmological model with $\Omega_\mathrm{M}=0.3$, $\Omega_\Lambda=0.7$ and $H_0=70$\,km/s/Mpc, in which $1\arcsec$ corresponds to $1.828$\,kpc at the redshift of the cluster.
Adjusting this cosmological model perturbs the inferred physical distances, and the absolute normalisation of inferred masses.

\section{Data} \label{sec:data}

\subsection{Pre-existing imaging}

\begin{figure*}
\begin{center}
\vspace{-1mm}
\includegraphics[width=0.95\textwidth]{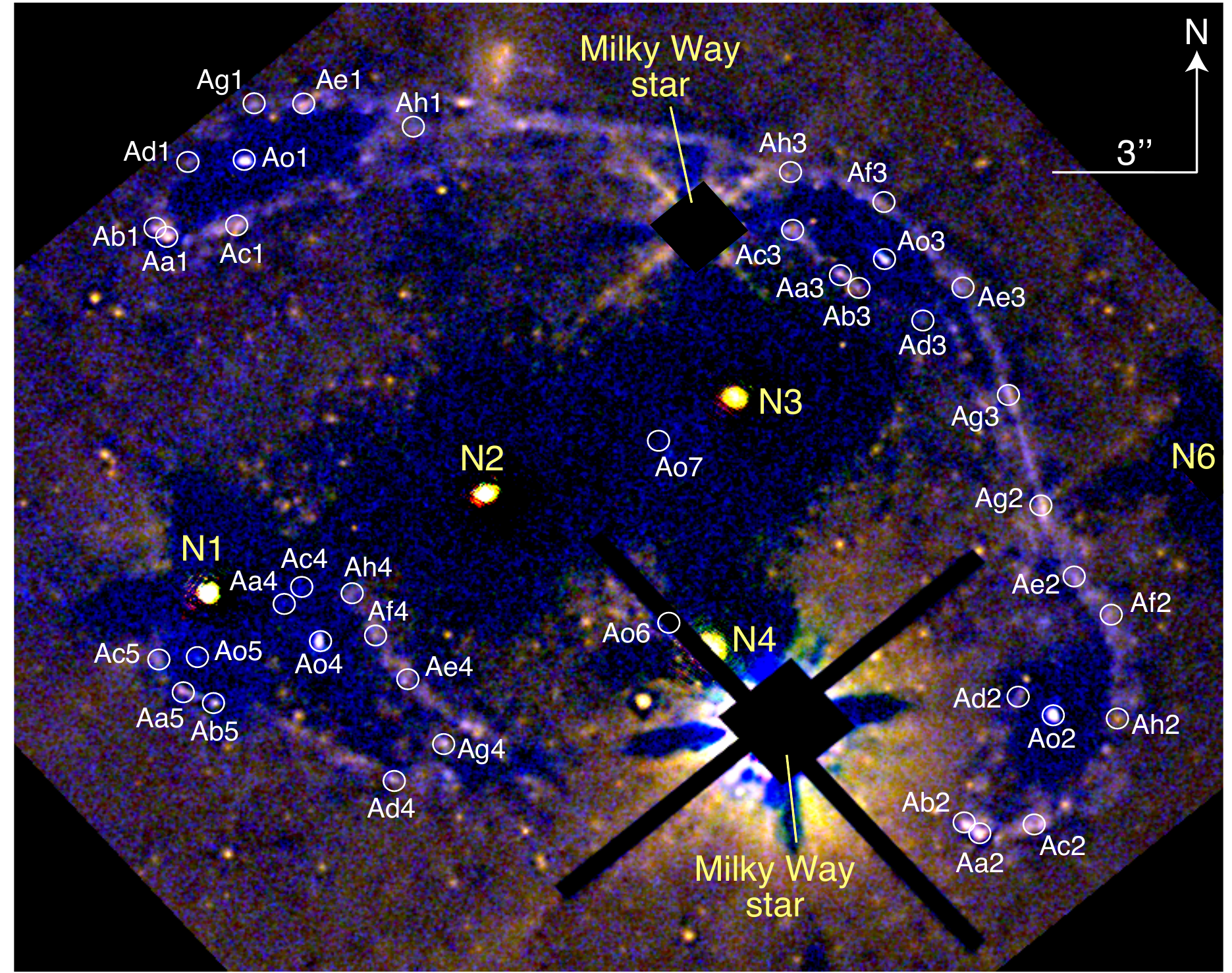}\\
\caption{{\sl Hubble Space Telescope} image of the core of Abell~3827 in F814W (red), F606W (green) and F336W (blue) bands.
Light from two foreground stars and five foreground galaxies (labelled in yellow) has been subtracted to reveal the background lens system.
The colour scale is linear.
Multiply-imaged components of the background spiral galaxy, identified either in this image or in ALMA/MUSE data are labelled in white. 
In our cosmological model, $3\arcsec=5.5$\,kpc at the redshift of the cluster.
} \label{fig:hst}
\end{center}
\end{figure*}

Broad-band imaging of Abell 3827 has been obtained by the 
Gemini telescope at optical wavelengths \citep{car10} and by 
the {\sl Hubble Space Telescope} ({\sl HST}; programme GO-12817) in the F336W (UV), F606W and F814W (optical) and F160W (IR) bands \citep{mas15}. 

This revealed four similarly-bright elliptical galaxies (N1--N4) within 10~kpc radius, and a background lensed spiral galaxy (with a red bulge and blue spiral arms), whose multiple images are threaded throughout the cluster core.
In this paper, we exclusively use the HST imaging. 
As described in \cite{tay17}, we reveal the background lensed galaxy by fitting and subtracting foreground emission from the five brightest cluster galaxies and two Milky Way stars using the {\sc MuSCADeT} method \citep{muscadet} (figure~\ref{fig:hst}).

\subsection{ALMA integral field spectroscopy}

\begin{figure*}
\begin{center}
\includegraphics[width=0.97\textwidth]{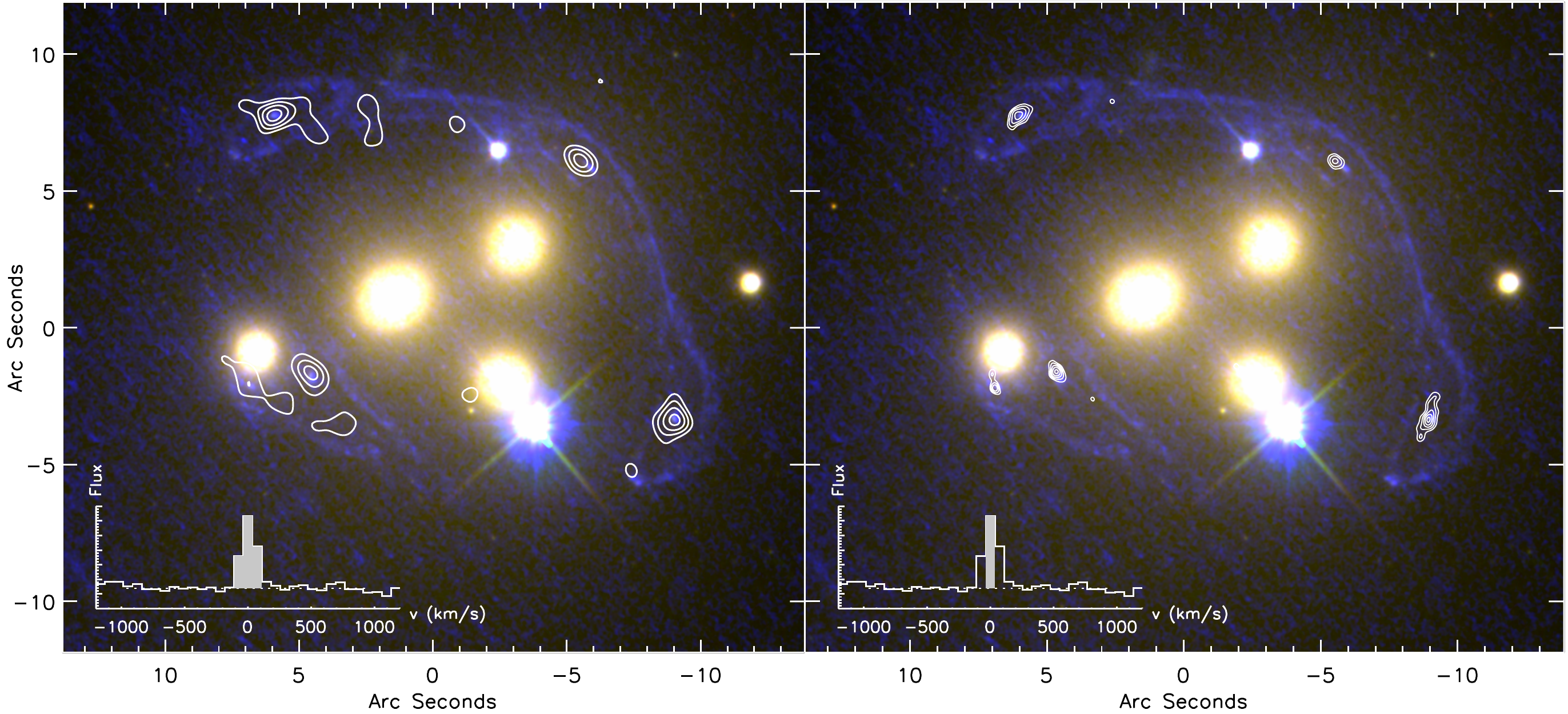}\\
\caption{
ALMA detection of CO(2-1) emission in the lensed spiral, as contours overlaid on the HST image from figure~\ref{fig:hst}, before foreground subtraction.
{\it Left:} CO(2-1) emission collapsed over $\pm100$\,km/s from the systemic redshift and $(u,v)$ tapered to a $0.8$\arcsec beam, to show the full emission.
The $1\sigma$ noise level is $0.15$\,mJy/beam, and contours show $3,4,5,6\sigma$.
{\it Right:} CO(2-1) emission from a single, central ALMA channel, at natural $0.47$\arcsec resolution, to identify multiple images of the source's bulge.
The $1\sigma$ noise level is $0.08$\,mJy/beam, and contours show $4,5,6,7\sigma$.
The inset spectra have a linear scale and include a dotted line at zero flux.
}
\label{fig:alma}
\end{center}
\end{figure*}

In October 2016, we obtained a 5.2 hour observation of Abell~3827 with the Atacama Large Millimetre Array (ALMA; programme 2016.1.01201.S).
The band 3 data sample frequencies $89.9$-$93.8$\,GHz and $101.8$-$105.6$\,GHz with spectral resolution $15.6$\,MHz ($47.8$\,km/s).
Observations were conducted with 44 12\,m antennae in the C40-6 configuration.
Flux and bandpass calibration were obtained from J2056$-$4714, and the phase calibrator was J2208$-$6325.

Data were reduced using \textsc{casa} software v4.7.2 % \textsc{Common Astronomy Software Application} (\textsc{casa}) 
\citep{mcmullin2007casa}. 
Spectral data cubes were created using the \textsc{clean} algorithm, with channel averaging and natural weighting to maximise sensitivity.
This yielded a synthesised beam of $\sim$$0.48\arcsec\times0.39\arcsec$, and a $1\sigma$ noise level of 0.08\,mJy/beam for each $31.3$\,MHz channel. 
In addition, to minimise potential extended flux being resolved out, we created a second spectral cube with a $(u,v)$ taper applied that yielded a synthesised beam of $\sim$\,0.87$''\times $\,0.82$''$ and $1\sigma$ noise level of 0.15\,mJy/beam.

The background $z=1.24$ galaxy is visible in emission from the $230.5$\,GHz CO(2-1) transition, redshifted to $102.8$\,GHz (figure~\ref{fig:alma}).
However, the emission is fainter than expected from an extrapolation of near-IR emission
(a somewhat indirect chain using {\sc [Oii]} emission to estimate star formation rate and hence far-infrared luminosity, then using the \citealt{sol05} relation to predict CO luminosity).
Our exposure time was therefore only just sufficient to detect spatial structure in the line emission; no continuum emission is detected beneath the foregrounds.

\subsection{VLT/MUSE integral field spectroscopy} \label{sec:IFUreduction}

\begin{figure*}
\begin{center}
\includegraphics[width=0.78\textwidth]{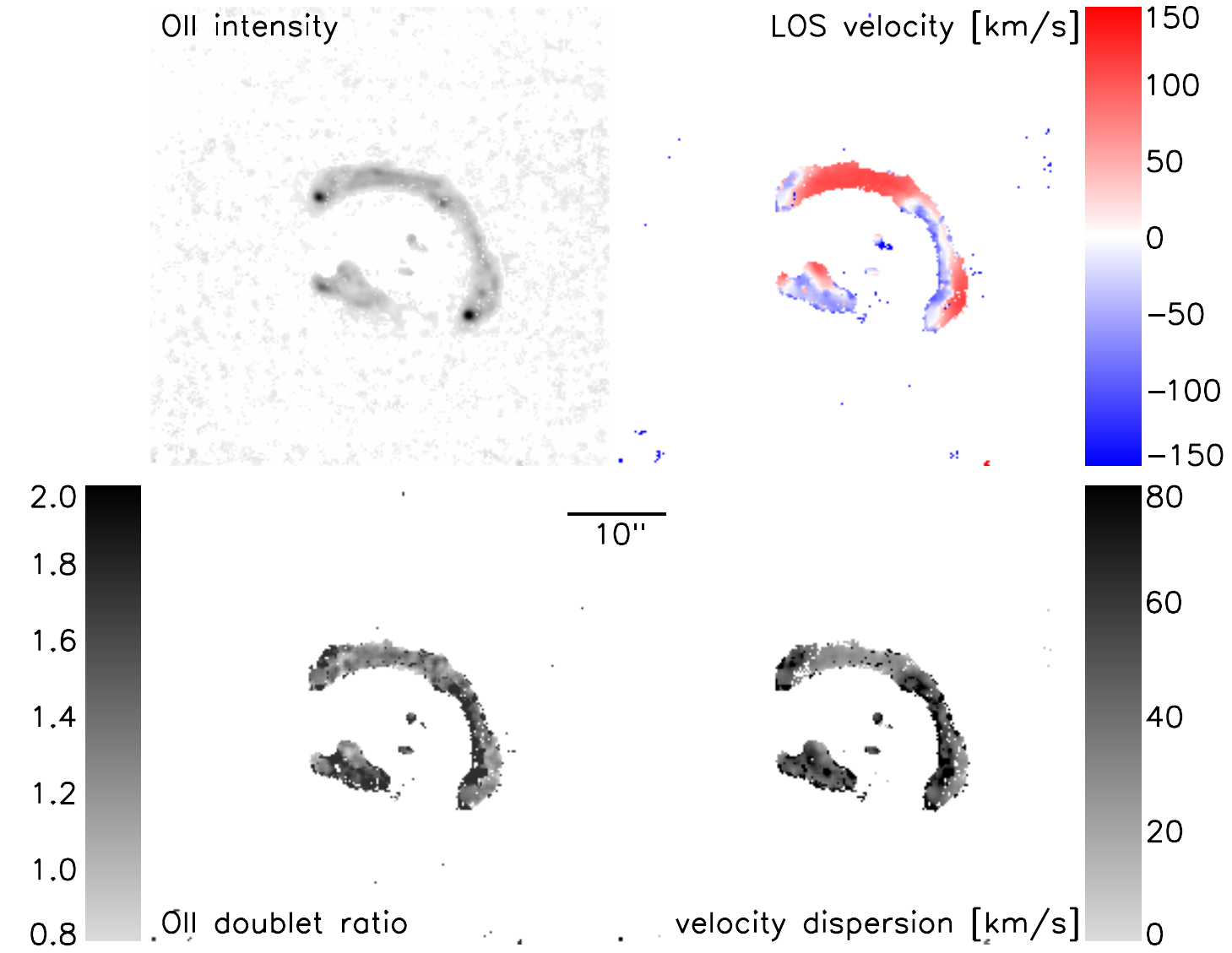}\\
\caption{MUSE data compressed into a 2D narrow-band image of [O{\sc ii}] doublet line emission from the lensed spiral galaxy, after subtraction of the foreground continuum emission {\it (top left)}.
To cross-identify regions of the galaxy in a way that is independent of the spatially varying lensing flux magnification, the other panels show parameters of a model fitted to the line doublet's spectral energy distribution in each spatial pixel where line emission is detected with S/N$>$$3$.
The parameters are the line's local line-of-sight velocity {\it (top right)}, the flux ratio between the doublet's two components {\it (bottom left)} and the spectral line width {\it (bottom right)}.
  }
\label{fig:muse}
\end{center}
\end{figure*}

In June 2016, we obtained a $4$\,hour integration of Abell~3827 using the Multi-Unit Spectroscopic Explorer (MUSE) Integral Field Unit (IFU) spectrograph \citep{muse} on the European Southern Observatory (ESO) Very Large Telescope (VLT).
We combined these data (programme 295.A-5018) with a pre-existing $1$\,hour exposure from programme 294.A-5014.
All the observations were obtained in dark time, with $V$-band seeing better than $0.7$\arcsec and good atmospheric transparency.  
The data sample wavelengths $475.0$-$935.1$\,nm with $0.125$\,nm binning and spectral resolution $R$$=$$4000$ at the red end.

Data were reduced using v1.0 of the {\sc esorex} pipeline, which extracts the spectra, applies wavelength and flat-field calibration, then forms the data cube.
Each hour on sky included $3\times$ $20$ minute exposures, dithered by $\sim$$10\arcsec$.
We aligned the individual exposures by registering the images of stars, then removed cosmic rays and pixel defects, and stacked the exposures using the {\sc exp\_combine} routine.
Flux calibration was achieved using ESO standard stars which were reduced in an identical manner.

The background galaxy is visible in emission from the [O{\sc ii}]\,$\lambda$3726.8,\,3729.2 line doublet, redshifted to 835.5\,nm.
In each spatial pixel, we model the spectrum of foreground continuum emission as a low-order polynomial either side of $835.5$\,nm.
We subtract this foreground emission, then integrate the remaining line flux as an [O{\sc ii}] narrow-band image (figure~\ref{fig:muse}).
We also use a two-Gaussian model to fit the [O{\sc ii}] doublet line ratio (3728.9/3726.2), line-of-sight velocity, and line width. 
Both components of the line are assumed to have the same width, and the measurement of spectral line width is corrected for instrumental broadening.

\section{Analysis} \label{sec:analysis}

\subsection{Strong lens image identifications}

The multiply-imaged background source is a spiral galaxy consisting of a red bulge `Ao' inside a blue ring of star formation knots `Aa'--`Ah'.
Its rotational support is apparent from the $\sim200$\,km/s velocity gradient apparent across the galaxy in the MUSE data (and present at low S/N in ALMA data, but not shown in figure~\ref{fig:alma}).

Features in the observed image have been variously identified as multiple versions of the background galaxy's bulge or star forming regions \citep{ws11,mas15,tay17}.
Many of these features are deeply embedded within the light from foreground galaxies.
After foreground subtraction using {\sc galfit} \citep{galfit}, and based on its apparent colour and morphology, \cite{mas15} identified a point-like source immediately south of galaxy N1 as the fifth (sorted by arrival time in the best-fit mass model) multiple image of knot Aa.

Our new data indicate that this identification was incorrect.
Our ALMA data show that the feature south of N1 is at the same systemic velocity and similar CO(2-1) flux as the source's central bulge.
The feature's line-of-sight velocity is also inconsistent with that of star formation knot Aa.
Our MUSE data also support a new interpretation that the feature is an additional image of the bulge, which we now call Ao5.
This identification of the source's central bulge implies that images of knots Aa and Ac must be further southeast.
The ALMA data is too low S/N to detect them, and the MUSE data have only barely sufficient angular resolution, but candidate features can be seen in HST imaging after our improved foreground subtraction using {\sc MuSCADeT} \citep{muscadet}.
These features were hidden behind the foreground emission from N1, and are fainter than the foreground cluster's many globular clusters.
Indeed, the chain of three or four sources between Ao4 and Ao5 appears to be an unfortunate alignment of foreground globular clusters, confusingly unrelated to the background source.

Building upon this new hypothesis, and incorporating additional features resolved by ALMA and ordered by MUSE, a new set of multiple-image identifications Ao and Aa--h become clear (Table~\ref{tab:multiple_image_IDs}).
This configuration of multiple image identifications was not amongst those considered in Appendix~B of \cite{mas15}.
We shall now demonstrate that this new configuration yields a model of the lens's mass distribution with higher Bayesian evidence and better consistency with observed lensed image positions.

\begin{table}
 \centering
  \caption{Locations of multiply-imaged components of the background spiral galaxy. 
  Images Ao$n$ are the bulge, and images A[a--h]$n$ are knots of star formation in the spiral arms.
  \label{tab:multiple_image_IDs}}
  \begin{tabular}{lrr}
  \hline
  \hline
Name\!\! & RA & Dec \\
\hline
Ao1 & $330.47479$ & $-59.94358$ \\
Ao2 & $330.46649$ & $-59.94665$ \\
Ao3 & $330.46828$ & $-59.94411$ \\
Ao4 & $330.47407$ & $-59.94623$ \\
Ao5 & $330.47529$ & $-59.94634$ \\
Ao6 & $330.47044$ & $-59.94614$ \\
Ao7 & $330.47054$ & $-59.94514$ \\
Aa1 & $330.47559$ & $-59.94400$ \\
Aa2 & $330.46725$ & $-59.94732$ \\
Aa3 & $330.46871$ & $-59.94421$ \\
Aa4 & $330.47443$ & $-59.94605$ \\
Aa5 & $330.47546$ & $-59.94652$ \\
Ab1 & $330.47571$ & $-59.94395$ \\
Ab2 & $330.46741$ & $-59.94726$ \\
Ab3 & $330.46852$ & $-59.94428$ \\
Ab5 & $330.47515$ & $-59.94658$ \\
Ac1 & $330.47487$ & $-59.94394$ \\
Ac2 & $330.46669$ & $-59.94726$ \\
Ac3 & $330.46920$ & $-59.94396$ \\
Ac4 & $330.47424$ & $-59.94596$ \\
Ac5 & $330.47571$ & $-59.94634$ \\
Ad1 & $330.47537$ & $-59.94359$ \\
Ad2 & $330.46685$ & $-59.94656$ \\
Ad3 & $330.46784$ & $-59.94446$ \\
Ad4 & $330.47327$ & $-59.94701$ \\
Ae1 & $330.47420$ & $-59.94327$ \\
Ae2 & $330.46627$ & $-59.94589$ \\
Ae3 & $330.46745$ & $-59.94428$ \\
Ae4 & $330.47315$ & $-59.94644$ \\
Af2 & $330.46589$ & $-59.94610$ \\
Af3 & $330.46826$ & $-59.94381$ \\
Af4 & $330.47348$ & $-59.94620$ \\
Ag1 & $330.47471$ & $-59.94327$ \\
Ag2 & $330.46661$ & $-59.94550$ \\
Ag3 & $330.46694$ & $-59.94488$ \\
Ag4 & $330.47276$ & $-59.94681$ \\
Ah1 & $330.47305$ & $-59.94340$ \\
Ah2 & $330.46583$ & $-59.94667$ \\
Ah3 & $330.46922$ & $-59.94364$ \\
Ah4 & $330.47372$ & $-59.94599$ \\
\hline
\hline
  \end{tabular}
\end{table}

\subsection{Mass model}

To ensure that we can draw robust conclusions, we use two independent algorithms to infer the mass distribution in the lens. 
Both have been tested in a blind analysis of strong lensing data for which the true mass distribution is known \citep{men17}.
First, we use {\sc lenstool} v6.8.1 \citep{lenstool}.
Its parametric mass models may not capture all the complexity of a real mass distribution, but it allows quantities of scientific interest (such as the position of dark matter) to be parameterised explicitly and to be fitted directly from data.
Second, we use {\sc grale} \citep{lie06}.
This `freeform' method possesses more flexibility to represent a real mass distribution and, by inferring unphysical distributions, to highlight errors in e.g.\ source image identification.
However, interpretation is later required to extract quantities of scientific interest.

\subsubsection{LENSTOOL}

Our {\sc lenstool} mass model consists of dark matter in one cluster-scale Pseudo-Isothermal Elliptical Mass Distribution (PIEMD; \citealt{limousin05,eliasdottir07}), plus the four bright galaxies' stellar and dark matter with respectively Hernquist and Pseudo-Isothermal Skewed Potential (PISP; \citealt{tay17}) distributions.
A PISP distribution reduces to a PIEMD if its skewness $s=0$.
We also fit a PISP component to the dark matter associated with faint member galaxy N6, but assume it has negligible stellar mass and skewness to reduce parameter space. 
Including mass associated with galaxies farther from the cluster core yields indistinguishable results but slows the analysis dramatically, so we omit them.
Finally, we allow an external shear \citep[e.g.][]{hog94}.

Parameters of the dark matter components are adjusted to reduce the rms of distances between the source galaxy's predicted and observed positions  in the image plane, $\langle$rms$\rangle$$_i$.
The parameters' posterior probability distribution function (PDF) is explored by a Markov Chain Monte Carlo (MCMC) iteration, with a constant proposal distribution after a burn-in phase ({\sc lenstool}'s {\sc runmode}=3) and priors identical to those in \cite{tay17}.
For example, the location and amount of each galaxy's dark matter is given a flat prior $2\arcsec$ either side of its stars. 
\cite{tay17} reported failed convergence of skewness parameters; this has been solved by a much longer Markov Chain that samples the PDF 100,000 times, and by ensuring that the skewness angle $\phi_s$ wraps far from any peak in the PDF.
We assume statistical uncertainty of 0.5\arcsec\ on the location of Ao6 and Ao7, which are detected only in ground-based data, and 0.15\arcsec\ on the location of every other image. 

Parameters of the stellar mass components are derived from {\sc Galfit} fits to flux in the F606W band, with the flux converted into mass via \citet{bc03} models, assuming a \citet{cha03} initial mass function, solar metallicity, and a single burst of star formation at redshift $z_{\mathrm f}\!=\!3$.
These parameters are fixed during the optimisation.

\subsubsection{GRALE}

Our {\sc grale} mass model incorporates a grid of approximately 1300 Plummer spheres \citep{plum11}
in a $50\arcsec\times50\arcsec$ region centered on (RA: $330.47043$, Dec: $-59.945804$).
An iterative procedure adaptively refines the grid in dense regions, and uses a genetic algorithm to adjust the mass in each Plummer sphere.
The genetic algorithm optimises the product of (a) the fractional degree of overlap between multiple images of the same source in the source plane and (b) a fitness measure penalising the presence of false counter-images in regions where they are not observed.

We run twenty mass reconstructions with different random seeds.
n total, this produces 26786 optimised Plummer spheres.
We average the inferred mass distributions; their rms provides an estimate of statistical error.

\section{Results} \label{sec:results}

\begin{figure}
\begin{center}
\includegraphics[trim = 8mm 72mm 8mm 37mm, clip, width=0.418\textwidth]{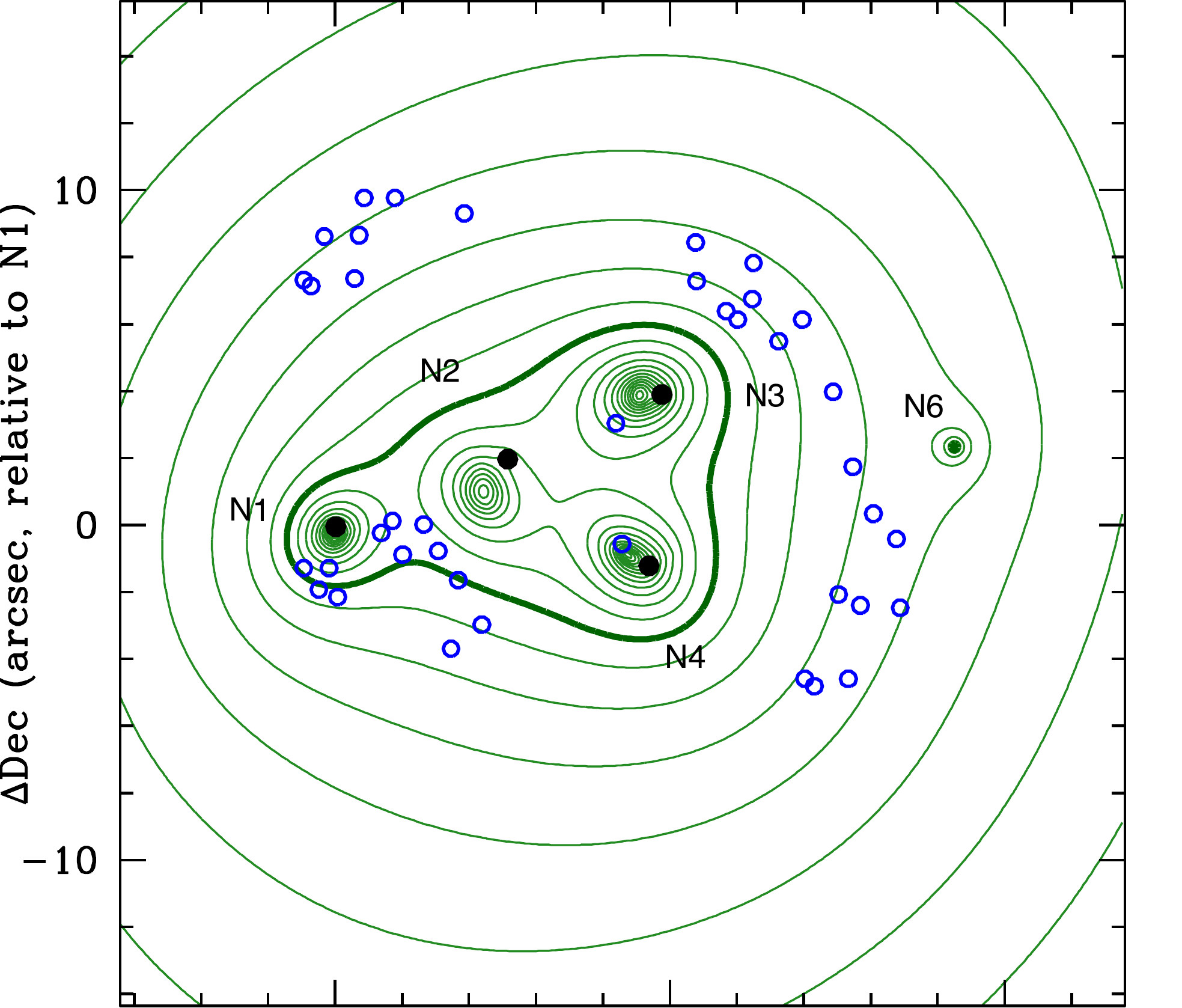}\\
\includegraphics[trim = 8mm 54mm 8mm 37mm, clip, width=0.418\textwidth]{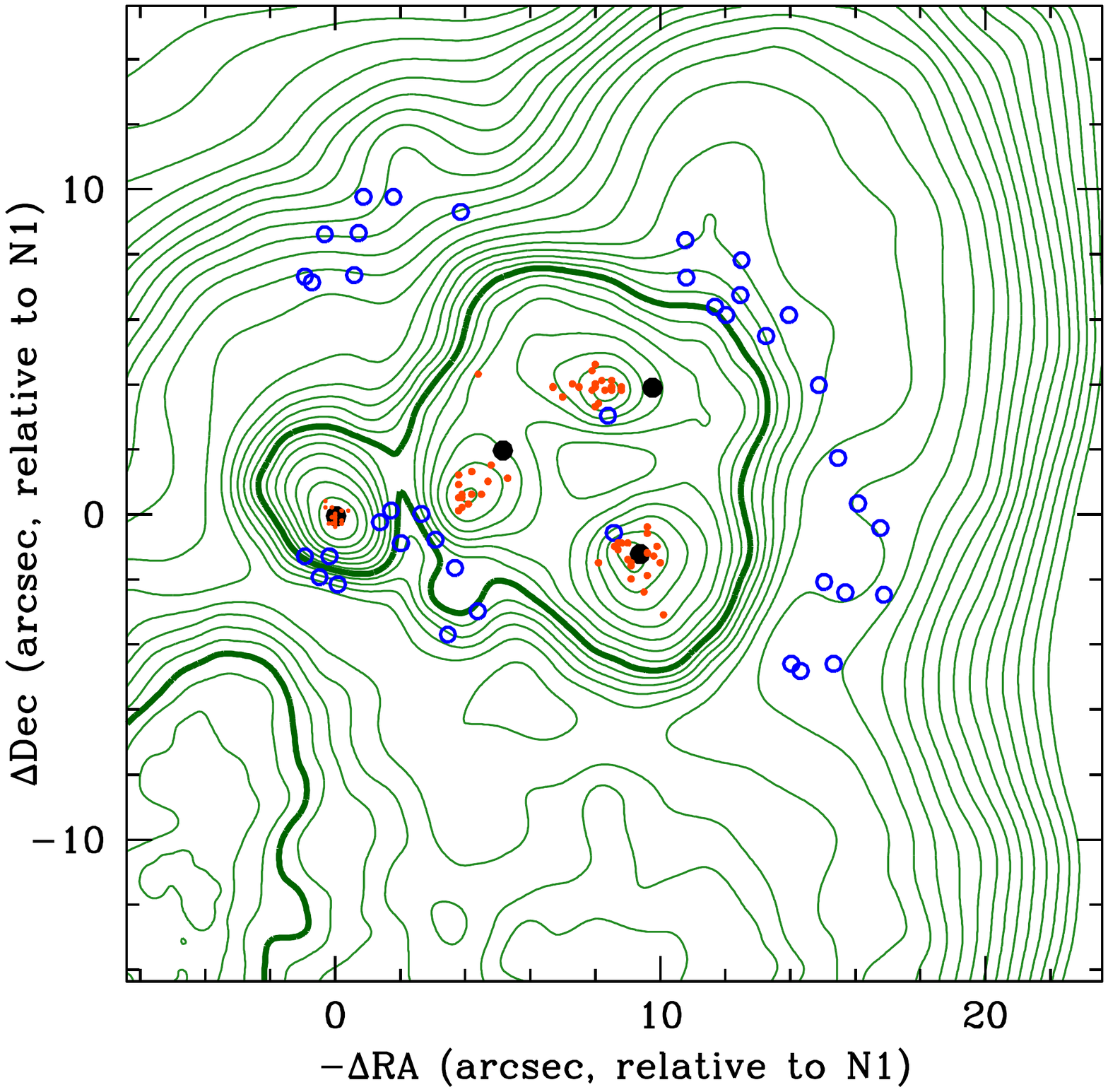}\\
\caption{{\it Top:} Map of total mass in the cluster core, reconstructed using {\sc lenstool}, and averaging over the posterior PDF.
Green contours show the projected mass density, spaced logarithmically by a factor 1.15; the thick contour shows convergence $\kappa\!=\!1$ for $z_{c\ell}\!=\!0.099$ and $z_\mathrm{A}\!=\!1.24$ ($\Sigma_\mathrm{crit}\!=\!1.03\,$g/cm$^2$). 
Blue circles show the lensed images.
Black dots show cluster ellipticals N1--N4. 
{\it Bottom:} Total mass, as in the top panel but reconstructed via {\sc grale}.
Red dots show local maxima in individual realisations of the mass map. 
}
\label{fig:massmaps}
\end{center}
\end{figure}

Inferred mass maps are presented in figure~\ref{fig:massmaps}.
Results from {\sc lenstool} and {\sc grale} are now more consistent with each other.
They also provide a better fit to the data than they were when assuming the source identifications of \cite{mas15} (whose {\sc lenstool} model had $\langle$rms$\rangle_i$=$0.26\arcsec$).
The new parameters of {\sc lenstool}'s best-fit model are presented in Table~\ref{tab:pots}.
This model achieves $\langle$rms$\rangle_i$=$0.13\arcsec$, or $\chi^2$=$31.7$ with $29$ degrees of freedom, likelihood log($\mathcal{L}$)=$59.9$, and Bayesian evidence log($\mathcal{B}$)=$-11.5$.

\begin{table*}
 \centering
  \caption{
Parameters of {\sc lenstool}'s best-fit mass model. 
Quantities in square brackets were fixed during optimisation. 
Errors on other quantities show 68\% statistical confidence limits, marginalising over uncertainty in all other parameters.
Stellar mass components are modelled as Hernquist profiles, with their mass, scale radius and ellipticity calculated from F606W broad-band emission.
Dark matter components are modelled as PIEMDs with a 1D velocity dispersion, core and cut radii, and ellipticity; or PISPs with an additional skewness.
Positions are given in arcseconds relative to (R.A.: $330.47518$, Dec.: $-59.945985$), except galaxies' dark matter components, which are relative to the position of their stars.
Angles are anticlockwise from West.
The external shear is ($1.47^{+0.97}_{-0.01}$)\%, at angle ($92^{+24}_{-94}$)$^\circ$.
  \label{tab:pots}}
  \begin{tabular}{lrr@{}lr@{}lr@{}lr@{}lr@{}lr@{}lr@{}lr@{}lr@{}lr@{}lr@{}lr@{}l}
  \hline
  \hline
 & & \multicolumn{2}{c}{$x\,$[\arcsec]} & \multicolumn{2}{c}{$y\,$[\arcsec]} & \multicolumn{2}{c}{Mass\,[$M_\odot$]} & \multicolumn{2}{c}{$r_\mathrm{sc}\,$[\arcsec]} & & & \multicolumn{2}{c}{\multirow{2}{*}{$\epsilon$~~~}} & \multicolumn{2}{c}{\multirow{2}{*}{$\phi_\epsilon\,$[$^\circ$]~}} & \multicolumn{2}{c}{\multirow{2}{*}{$s$}} & \multicolumn{2}{c}{\multirow{2}{*}{$\phi_s\,$[$^\circ$]}}\\ 
 & & \multicolumn{2}{c}{$\Delta x\,$[\arcsec]} & \multicolumn{2}{c}{$\Delta y\,$[\arcsec]} & \multicolumn{2}{c}{$\sigma_\mathrm{v}\,$[km/s]} & \multicolumn{2}{c}{$r_\mathrm{core}\,[\arcsec]$} & \multicolumn{2}{c}{\!\!$r_\mathrm{cut}\,[\arcsec]$\!\!}
 \\
\hline
N1 & stars & [$ 0.00$ & ] & [$ 0.00$ & ] & \multicolumn{2}{c}{[$ 1.00\times10^{11}$]} & \multicolumn{2}{c}{[$0.53$]} & &  & [$0.12$ & ] & [$61$ & ] \\
\multicolumn{2}{r}{dark matter} & $0.09$ & $^{+0.10}_{-0.09}$ & $-0.28$ & $^{+0.13}_{-0.12}$ & ~~~$166$ & $^{+9}_{-11}$ & \multicolumn{2}{c}{[$0.10$]} & \multicolumn{2}{c}{[$40$]} & $0.12$ & $^{+0.25}_{+0.00}$ & $56$ & $^{+52}_{-18}$ & $0.14$ & $^{+0.08}_{-0.28}$ & $102$ & $^{+12}_{-106}$ \vspace{2mm} \\
N2 & stars & [$ 5.13$ & ] & [$ 2.00$ & ] & \multicolumn{2}{c}{[$ 2.47\times10^{11}$]} & \multicolumn{2}{c}{[$0.79$]} & &  & [$0.17$ & ] & [$39$ & ] \\
\multicolumn{2}{r}{dark matter} & $-0.81$ & $^{+0.19}_{-0.20}$ & $-0.59$ & $^{+0.33}_{-0.29}$ & ~~~$170$ & $^{+13}_{-18}$ & \multicolumn{2}{c}{[$0.10$]} & \multicolumn{2}{c}{[$40$]} & $0.38$ & $^{+0.01}_{-0.25}$ & $129$ & $^{+16}_{-22}$ & $0.10$ & $^{+0.09}_{-0.15}$ & $41$ & $^{+94}_{-23}$ \vspace{2mm} \\
N3 & stars & [$ 9.75$ & ] & [$ 3.93$ & ] & \multicolumn{2}{c}{[$ 2.76\times10^{11}$]} & \multicolumn{2}{c}{[$0.33$]} & &  & [$0.05$ & ] & [$31$ & ] \\
\multicolumn{2}{r}{dark matter} & $-0.57$ & $^{+0.14}_{-0.14}$ & $0.08$ & $^{+0.24}_{-0.16}$ & ~~~$214$ & $^{+6}_{-14}$ & \multicolumn{2}{c}{[$0.10$]} & \multicolumn{2}{c}{[$40$]} & $0.14$ & $^{+0.07}_{-0.08}$ & $14$ & $^{+18}_{-8}$ & $-0.09$ & $^{+0.09}_{-0.07}$ & $41$ & $^{+67}_{-27}$ \vspace{2mm} \\
N4 & stars & [$ 9.32$ & ] & [$-1.12$ & ] & \multicolumn{2}{c}{[$ 2.06\times10^{11}$]} & \multicolumn{2}{c}{[$1.37$]} & &  & [$0.39$ & ] & [$127$ & ] \\
\multicolumn{2}{r}{dark matter} & $-0.54$ & $^{+0.34}_{-0.11}$ & $0.40$ & $^{+0.09}_{-0.20}$ & ~~~$206$ & $^{+7}_{-15}$ & \multicolumn{2}{c}{[$0.10$]} & \multicolumn{2}{c}{[$40$]} & $0.32$ & $^{+0.33}_{+0.00}$ & $144$ & $^{+12}_{-65}$ & $0.12$ & $^{+0.11}_{-0.12}$ & $104$ & $^{+53}_{-58}$ \vspace{2mm} \\
N6 & stars & [$18.60$ & ] & [$2.43$ & ] & \multicolumn{2}{c}{[0]} \\
\multicolumn{2}{r}{dark matter} & [$ 0.00$ & ] & [$ 0.00$ & ] & ~~~$61$ & $^{+13}_{-27}$ & \multicolumn{2}{c}{[$0.10$]} & \multicolumn{2}{c}{[$40$]} & [$0.00$ & ] & [$0$ & ] & [$0$ & ] & [$0$ & ] \vspace{2mm} \\
Cluster & dm & $8.61$ & $^{+0.89}_{-0.90}$ & $-0.28$ & $^{+1.04}_{-0.79}$ & ~~~$842$ & $^{+77}_{-89}$ & ~~$30$ & $^{+5}_{-7}$ & \multicolumn{2}{c}{[$1000$]} & $0.50$ & $^{+0.07}_{-0.15}$ & $62$ & $^{+2}_{-2}$ & [$0$ & ] & [$0$ & ] \\
\hline\hline
  \end{tabular}
\end{table*}

Central images Ao6 and Ao7 have split {\sc grale}'s previous reconstruction of a bimodal cluster (consisting of N1 plus everything else) into four distinct mass concentrations around each galaxy. 
There is no reason for the genetic algorithm to prefer either, yet the new model is more physical.
Adding the central images also creates a prediction (from both {\sc lenstool} and {\sc grale}) for a diffuse trail of source emission southwest of Ag4, including counter-images of Ab, Ad, and Ag.
These are possibly demagnified and observed, but the whole area is unclear in HST imaging because of the bright foreground star and confusion with globular clusters.
Both models predict demagnified images of the star formation knots tightly packed around Ao6, and loosely packed around Ao7, as are visible in MUSE data but with insufficient confidence to be used as input constraints.

Our {\sc lenstool} analysis suggests a $\sim2\sigma$ statistical significance for the offset of N1.
However, the absolute value of the offset is far smaller than in \cite{mas15}, and its significance disappears entirely when combining with our {\sc grale} analysis and allowing for systematic, model-induced biases of up to $\sim$$0.21\arcsec$ for this configuration of lenses \citep{mas15}. 
The mass peak reconstructed by {\sc grale} outside the cluster core imposes an external shear near N1 consistent with that fitted by {\sc lenstool}.

Statistical errors on the position of dark matter associated with N2, N3 and N4 are tightened by our new detection of central images Ao6 and Ao7.
They would be dramatically improved if more of the source galaxy's structure could be seen in the central images (e.g.\ with deeper ALMA data).
However, the position of N2's dark matter shows a large scatter in our current {\sc grale} analysis, and can change in a {\sc lenstool} analysis if the prior is adjusted on the position of the cluster-scale halo.
In the MCMC chain of our {\sc lenstool} analysis, the positions of N3 and N4 are degenerate with each other.
Furthermore, we have an {\it a priori} expectation that only N1 is sufficiently close to space-resolution lensed images to be constrained with kiloparsec accuracy (even when the lens identifications are unambiguous \citealt{har16}, and they may still not all be correct here).

The inferred location of the dark matter associated with each galaxy N1--N4 appears consistent with the location of its stars.
Deeper ALMA or {\sl HST} observations would clarify the status of N2, N3 and N4. 
However, given various parameter degeneracies in our current analysis, and the potential for systematic errors at a level comparable to their offsets, we cannot here conclude that any offset is physically significant.

The total mass of the dark matter components of galaxies N1--N4 is formally $1.47^{+0.16}_{-0.19}$, $1.54^{+0.24}_{-0.31}$, $2.44^{+0.14}_{-0.31}$ and $2.26^{+0.16}_{-0.32}\times10^{12}\,M_\odot$, and that of the cluster-scale halo is $2.79^{+0.53}_{-0.56}\times10^{14}\,M_\odot$ \citep[see equation 10 of][]{limousin05}.
However, these calculations depend approximately linearly on our unconstrained choice of $r_{\mathrm{cut}}$.

\begin{figure}
\includegraphics[width=0.475\textwidth]{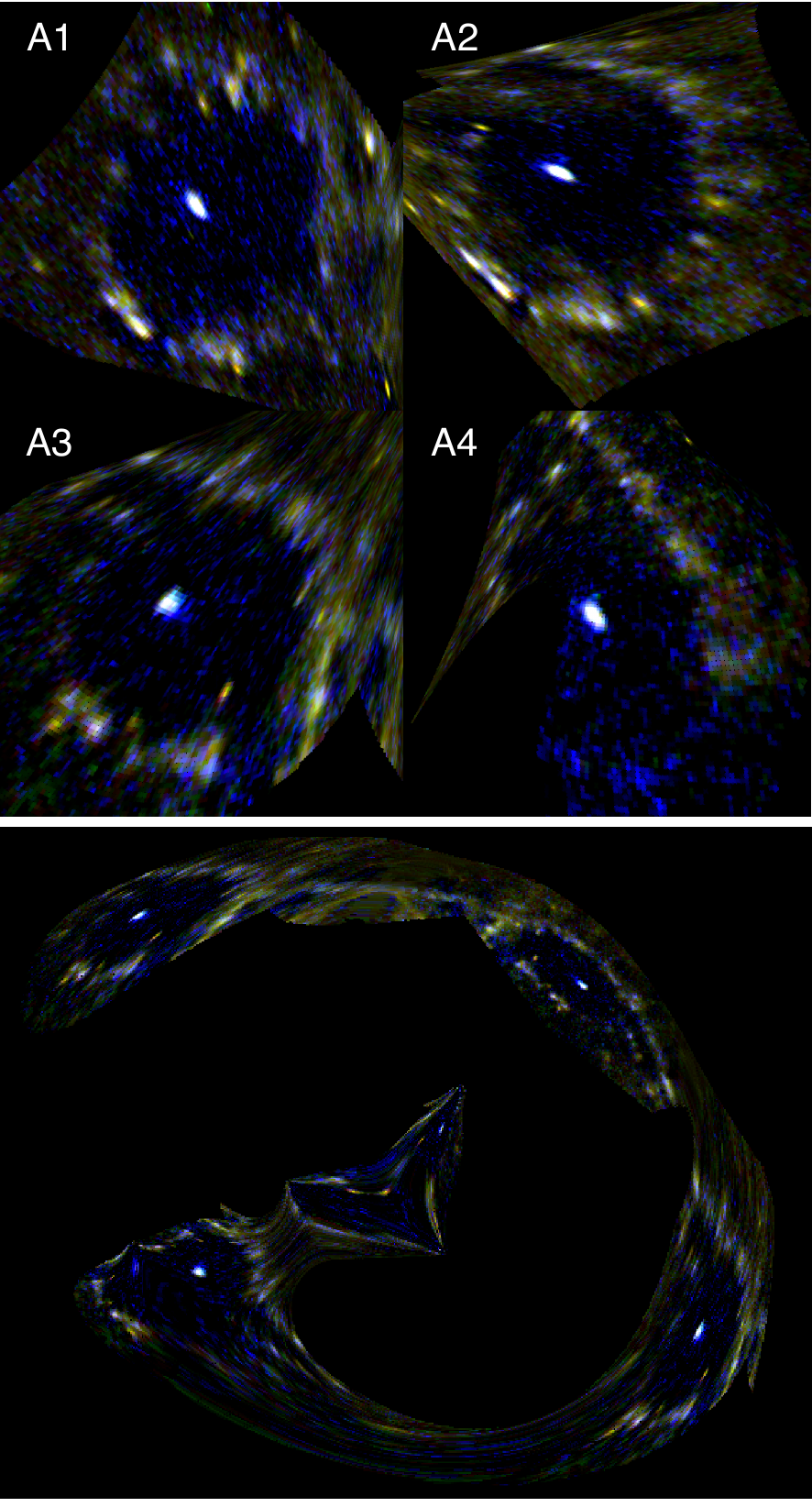}\\
\caption{{\it Top:} De-lensed images of the $z=1.24$ source galaxy, after foreground subtraction as in figure~\ref{fig:hst} and assuming the best-fit lens model produced by {\sc lenstool} in table~\ref{tab:pots}.
Each panel is $1.5\arcsec\times1.5\arcsec$, and centered on (RA=$22$h\,$01\arcmin$\,$53\farcs0$, Dec=$-59^\circ$\,$56\arcmin$\,$44\arcsec$).
Results from {\sc grale} are similar.
{\it Bottom:} Re-lensed version of the above realisation of source A3, the centre of the triple. 
The predicted brightness of the central images Ao6 and Ao7 changes sightly if other versions of the source are re-lensed.}
\label{fig:sources}
\end{figure}

De-lensed images of the background galaxy, assuming the best-fit {\sc lenstool} model, are presented in figure~\ref{fig:sources}; results from {\sc grale} are similar.
It is a ring galaxy reminiscent of the $z=1.67$ lensed source in Zwicky cluster Cl0024+1654 \citep{col96,jon10}.
Its central component is by far the brightest in CO(2-1) emission.
A large reservoir of dusty, molecular gas in a galaxy's bulge would be unusual at $z=0$, but not at $z=1.24$, when bulges are still forming stars.
Assuming {\sc lenstool}'s best-fit mass model, the luminosity-weighted amplification of its {[\sc Oii]} emission is $\mu=144$, summing over all the images.
Taking into account this amplification, its apparent {[\sc Oii]} luminosity implies an total star-formation rate of $\sim1\,M_\odot$/yr, using the \cite{ken98} calibration and a \cite{cha03} initial mass function.
Canonical dust extinction of about $A_V$$\sim$$1$\ magnitude could raise this by a factor $2$--$3$. 
Other than its role in gravitational lensing due to its location behind a cluster, it is not an intrinsically unusual galaxy.

\section{Conclusions} \label{sec:conc}

Previous studies of galaxy cluster Abell~3827 \citep{ws11,mas15} imaging suggested that the dark matter associated with at least one of its galaxies is offset from its stars.
This is predicted by simulations of self-interacting dark matter in which the exchange particle is light \citep{har14,kah14,rob17b}.
Prompted by this potentially exciting result, further simulations \citep{kah15}
suggested that the offset could be observable in (rare) systems where a massive galaxy intersects a cluster's Einstein radius, and its 3D motion happens to be near the plane of the sky. 
A strongly lensing merger between two field galaxies has shown a similar offset \citep{shu16}.

In this paper, ALMA has proved an exceptional tool to identify background lensed images, with high spatial resolution at wavelengths where foreground galaxy clusters are virtually transparent.
Whilst there is no guarantee that we have perfected the source identifications in Abell\,3827, it is now possible to construct lens models with residuals that are consistent with noise, and robust between very different modelling approaches.
The consistency between parametric and non-parametric lens models lends confidence to the conclusions. 
Indeed, both ALMA data and deviations from physically expected mass distributions in a free-form mass reconstruction could be a powerful discriminator between future source identifications.

Our new analysis shows that there is no statistically significant offset between galaxies and their dark matter in Abell~3827, projected onto the plane of the sky.
Galaxy N1 is best constrained. Assuming statistical errors only, its offset in our {\sc lenstool} model is 
$0.29^{+0.12}_{-0.13}$\,arcseconds or
$0.54^{+0.22}_{-0.23}$\,kpc.
Following \cite{kah15}'s reasoning that any offset requires dark matter self-interactions to balance a gravitational restoring force that can be calculated, this measurement implies an interaction cross-section 
$(\sigma/m)\cos{(i)}=0.68^{+0.28}_{-0.29}$\,cm${^2}$/g, 
where $i$ is the inclination of the galaxy's 3D motion with respect to the plane of the sky.
That this angle is unknown makes it difficult to infer an upper limit on $\sigma/m$ from this system without further information.

Nonetheless, the unusual configuration of Abell~3827, with four bright central galaxies and a background spiral galaxy with complex morphology is multiply-imaged between them, makes it still interesting for studies of dark matter dynamics.
Regardless of possible particle interactions, as a galaxy enters a cluster, its dark matter halo is gradually stripped via tidal gravitational forces. 
Simulations disagree about the timescale and the orbits on which dark matter stripping occurs in the inner tens of kiloparsecs \citep{die07,pen08,wet09,bah12}, but this dissipation is a key ingredient in semi-analytic models of galaxy formation \citep[e.g.][]{dar10}.
Observations of dark matter mass loss in galaxies entering a galaxy cluster from the field \citep{man06,lim07,lim12,par07,nat09,gil13,nie17} have never been followed inside $\sim$$1$\,Mpc, and measurements of strong lensing clusters with multiple central galaxies, like those in table~\ref{tab:pots}, could constrain this for the first time.

\section*{Acknowledgments}

The authors are grateful for helpful conversations with Jean-Paul Kneib, Subir Sarkar, and Felix Kahlhoefer.
RM and TDK are supported by Royal Society University Research Fellowships.
Durham authors were also supported by the UK Science and Technology Facilities Council (grant numbers ST/P000541/1, ST/H005234/1, ST/I001573/1 and ST/N001494/1). 
JL acknowledges the computational resources and services provided by the VSC (Flemish Supercomputer Center), funded by the Research Foundation Flanders (FWO) and the Flemish Government, department EWI.
LLRW would like to acknowledge the computational resources of the Minnesota Supercomputing Institute.
JM has received funding from the European Union's FP7 and Horizon 2020 research and innovation programmes under Marie Sk\l{}odowska-Curie grant agreement numbers 627288 and 664931.

\noindent {\it Facilities:}
This paper uses ALMA observations ADS/ JAO.ALMA\#2016.1.01201.S. ALMA is a partnership of ESO (representing its member states), NSF (USA) and NINS (Japan), together with NRC (Canada), NSC and ASIAA (Taiwan), and KASI (Republic of Korea), in cooperation with the Republic of Chile. The Joint ALMA Observatory is operated by ESO, AUI/NRAO and NAOJ.
This paper uses data from observations made with ESO Telescopes at the La Silla Paranal Observatory under programmes 093.A-0237 and 294.A-5014. We thank the Director General for granting discretionary time, and Paranal Science Operations for running the observations.
This paper uses data from observations GO-12817 with the NASA/ESA {\sl Hubble Space Telescope}, obtained at the Space Telescope Science Institute, which is operated by AURA Inc, under NASA contract NAS 5-26555.
All data are available from the telescopes' archives. 
This paper used the DiRAC Data Centric system at Durham University, operated by the Institute for Computational Cosmology on behalf of the STFC DiRAC HPC Facility (\url{www.dirac.ac.uk}). This equipment was funded by BIS National E-infrastructure capital grant ST/K00042X/1, STFC capital grant ST/H008519/1, and STFC DiRAC Operations grant ST/K003267/1 and Durham University. DiRAC is part of the UK National e-Infrastructure.

\bsp
\label{lastpage}

\end{document}